\documentclass[prd,twocolumn,showpacs,letterpaper,nofootinbib]{revtex4}
\usepackage{amsmath}

\newcommand{\be}{\begin{equation}}
\newcommand{\ee}{\end{equation}}
\newcommand{\bea}{\begin{eqnarray}}
\newcommand{\eea}{\end{eqnarray}}

\begin{document}

\title{Strings at future singularities.}
\author{Adam Balcerzak}
\email{abalcerz@sus.univ.szczecin.pl}
\author{Mariusz P. D\c{a}browski}
\email{mpdabfz@sus.univ.szczecin.pl}
\affiliation{Institute of Physics, University of Szczecin,
Wielkopolska 15, 70-451 Szczecin, Poland.}

\date{\today}

\begin{abstract}
We discuss the behaviour of strings propagating in spacetimes which
allow future singularities of either a sudden future
or a Big-Rip type. We show that in general the invariant string size remains
finite at sudden future singularities while it grows to infinity
at a Big-Rip. This claim is based on the discussion of both the
tensile and null strings. In conclusion, strings may survive a sudden
future singularity, but not a Big-Rip where they are infinitely stretched.
\end{abstract}


\maketitle

According to the claim of Ref. \cite{lazkoz04} sudden future
singularities (SFS) \cite{barrow04,lake,barrow042,sfs1} are not strong singularities
in the sense of Tipler and Kr\'olak \cite{krolak}. In particular, geodesic equations
do not feel anything special on the approach to SFS (there is no geodesic incompletness)
and the only sign of such singularities may be experienced by the extended objects which may
feel infinite tidal forces. On the other hand, Big-Rip
(BR) \cite{caldwell, phantom} singularities are the strong ones and, according to Ref.
\cite{lazkoz04}, they are felt by geodesics and also lead to the
destruction of structures.

In this context we discuss explicitly the behaviour of extended
objects such as fundamental strings \cite{vesan} at the future
singularities of either BR or SFS type. We will discuss a
possibility for a string to cross a SFS with its size conserved.
We shall analyze a possibility to blow-up a string at BR and to
collapse a string to a point at SFS.

Firstly, it is interesting to remark that at the Schwarzschild
horizon $r=r_s$ the metric is singular while the curvature
invariants $R, R_{\mu\nu}R^{\mu\nu}, R_{\mu\nu\rho\sigma}R^{\mu\nu\rho\sigma}$ are
finite. The SFS is different from the respect that the metric is
nonsingular at sudden future singularity for $t=t_s$,
but all the curvature invariants $R, R_{\mu\nu}R^{\mu\nu}, R_{\mu\nu\rho\sigma}R^{\mu\nu\rho\sigma}$
have a blow-up due to the blow-up of the Riemann tensor as a
consequence of the divergence of the second derivative of the
scale factor. In fact, the geodesic deviation equation feel
SFS due to the divergence of the Riemann tensor.

According to string theory a free string which propagates in a curved spacetime sweeps
out a world--sheet (a two-dimensional surface) in contrast to a point
particle whose history is a world-line. The world--sheet action for
such a string in the conformal gauge is given by \cite{green}
\begin{equation}
\label{curvesheet}
S= -\frac{T}{2} \int d\tau d\sigma
\eta^{ab} g_{\mu\nu} \partial_{a} X^{\mu} \partial_{b}
X^{\nu}   ,
\end{equation}
where $T=1/2\pi\alpha'$ is the string tension,
$\alpha'$ the Regge slope, $\tau$ and $\sigma$
are the (spacelike and timelike, respectively) string coordinates,
$\eta^{ab}$ is a 2-dimensional world--sheet (flat) metric $(a,b = 0,1)$,
$X^{\mu}(\tau,\sigma)$
($\mu, \nu = 0, 1, 2, 3)$ are the coordinates of the string world--sheet
in a 4-dimensional spacetime with metric
$g_{\mu\nu}$.

The equations of motion and the constraints for the action (\ref{curvesheet})
are \cite{green}
\begin{eqnarray}
\ddot{X}^{\mu}+\Gamma^{\mu}_{\nu\rho}\dot{X}^{\nu}\dot{X}^{\rho} & = &
\lambda ({X''}^{\mu}+\Gamma^{\mu}_{\nu\rho}{X'}^{\nu}{X'}^{\rho})\label{ruch1} \:, \\
g_{\mu\nu}\dot{X}^{\mu}\dot{X}^{\nu} & = &
-\lambda g_{\mu\nu}{X'}^{\mu}{X'}^{\nu} \label{ruch2} ,\\
g_{\mu\nu}\dot{X}^{\mu} {X'}^{\nu} & = & 0 \label{ruch3},
\end{eqnarray}
where: $(...)^{.}\equiv\frac{\partial}{\partial\tau}$,
$(...)'\equiv\frac{\partial}{\partial\sigma}$, and $\lambda = 1$ for tensile strings,
while $\lambda =0$ for tensionless/null strings \cite{null,dablar,iza02}. The meaning of the
constraints (\ref{ruch2}) and (\ref{ruch3})
for a null string is as follows. The first one shows that the
string moves with the speed of light and the second says that the
velocity vector of an element of string is perpendicular to this
element.

An important characteristic for strings
is their invariant size for closed strings defined by \cite{green}
\begin{equation}
\label{sizesig}
S(\tau)=\int_0^{2\pi}\;\sqrt{g_{\mu \nu}X^{\prime
\mu}X^{\prime\nu}}\;d\sigma~.
\end{equation}

Expressing general spacetime coordinates as: $X^{0}=t(\tau,\sigma)$, $X^{1}=r(\tau,\sigma)$,
$X^{2}=\theta(\tau,\sigma)$, $X^{3}=\varphi(\tau,\sigma)$
the equations of motion (\ref{ruch1}) for a string propagating in a Friedmann spacetime are:
\begin{eqnarray}
\label{rneom1} \ddot{t} & - & \lambda
{t''}+ aa_{,t} \left[f^2(r) (\dot{t}^2 - \lambda {t'}^2) +
r^2 (\dot{\theta}^2 - \lambda {\theta'}^2) \right. \nonumber \\
&&\left. + r^2 sin^2{\theta} (\dot{\varphi}^2 -
\lambda {\varphi'}^2 ) \right]=0~, \\
\label{rneom2} \ddot{r} & - & \lambda
{r''}+ 2\frac{a_{,t}}{a} ( \dot{t} \dot{r} - \lambda t' r' )
+ kr f^2(r) (\dot{r}^2 - \lambda {r'}^2) \\
& - & \frac{r}{f^2(r)} ( \dot{\theta}^2 - \lambda {\theta'}^2 ) -
\frac{r \sin^2{\theta}}{f^2(r)} (\dot{\varphi}^2 - \lambda
{\varphi'}^2 ) = 0 ~, \nonumber \\
\label{rneom3}
\ddot{\theta} & - & \lambda
{\theta''}+ 2\frac{a_{,t}}{a} ( \dot{t} \dot{\theta} - \lambda t' \theta' )
+ \frac{2}{r} (\dot{r} \dot{\theta} - \lambda r' \theta' )
\nonumber \\
& - & \sin{\theta} \cos{\theta} (\dot{\varphi}^2 - \lambda {\varphi'}^2 ) = 0 \:,\\
\label{rneom4} \ddot{\varphi} & - &
\lambda \varphi''+ 2\frac{a_{,t}}{a} ( \dot{t} \dot{\varphi} - \lambda t' {\varphi}')
+ \frac{2}{r} ( \dot{r}\dot{\varphi}-\lambda r'\varphi' )
\nonumber \\
&+& 2\cot\theta(\dot{\theta}\dot{\varphi}-\varepsilon^{2}\theta'\varphi')=0\:,
\end{eqnarray}
whereas the constraints (\ref{ruch2})-(\ref{ruch3}) are given by
\begin{eqnarray}
\label{constr1}
&-& \dot{t}^2 + \lambda {t'}^2 + a^2(t) f^2(r) ( \dot{r}^2 - \lambda {r'}^2 )
+ a^2(t) r^2 ( \dot{\theta}^2 - \lambda {\theta'}^2 )
\nonumber \\
&+& a^2(t) r^2 \sin^2{\theta} (\dot{\varphi}^2 - \lambda
{\varphi'}^2 ) = 0\:,\\
\label{constr2}
&-& \dot{t}t'- a^2(t) f^2(r) \dot{r}r' + a^2(t) r^2 \dot{\theta}\theta'
\nonumber \\
&+& a^2(t)r^2 \sin^{2}\theta\dot{\varphi}\varphi'=0\:.
\end{eqnarray}
Here $a(t)$ is the scale factor and $f^2(r) = 1/(1-kr^2)$ with $k=0,\pm 1$
- the curvature index. The invariant string size (\ref{sizesig}) is
\bea
&& S(\tau) \\
&=& \int_0^{2\pi} \sqrt{-{t'}^2 + a^2 f^2 {r'}^2 + a^2 r^2
{\theta'}^2 + a^2 r^2 \sin^2{\theta} {\varphi'}^2 } d\sigma
\nonumber \ .
\eea
One of the simplest string configurations is
a circular string given by the ansatz \cite{dablar,iza02}:
\begin{eqnarray}
\label{circular}
t=t(\tau), r=r(\tau), \theta=\theta(\tau), \varphi=\sigma,
\end{eqnarray}
It gives (\ref{rneom1})-(\ref{rneom4}) as
\bea
\label{eom1}
\ddot{t} &+& aa_{,t} [ f^2 \dot{t}^2 + r^2 \dot{\theta}^2 - \lambda
r^2 \sin^2{\theta} ] = 0~, \\
\label{eom2}
\ddot{r} &+&  2\frac{a_{,t}}{a} \dot{t} \dot{r} + krf^2 \dot{r}^2
- \frac{r}{f^2} \dot{\theta}^2 + \lambda
\frac{r\sin^2{\theta}}{f^2} = 0~,\\
\label{eom3}
\ddot{\theta} &+& 2\frac{a_{,t}}{a} \dot{t} \dot{\theta} +
\frac{\dot{r}}{r} \dot{\theta} + \lambda \sin{\theta} \cos{\theta}
= 0~,\\
\label{eom4}
\ddot{\varphi} &=& 0~,
\eea
and the constraint (\ref{constr2}) is fulfilled automatically, while (\ref{constr1})
reads as
\bea
- \dot{t}^2 &+& a^2 f^2 \dot{r}^2
+ a^2 r^2 \dot{\theta}^2 - \lambda a^2 r^2 \sin^2{\theta}
= 0   ~.
\eea
The invariant string size now reads as
\bea
\label{circsize}
S(\tau) &=& 2\pi a(t(\tau)) r(\tau) \sin{\theta(\tau)}~.
\eea

Let us first discuss briefly the tensile strings and then the null
strings. The flat Friedmann Universe admits a different circular ansatz
given by Cartesian coordinates \cite{FRWinigo}
\begin{equation}
\label{cartcircular}
t=t(\tau), X=R(\tau)\cos{\sigma}, Y=R(\tau)\sin{\sigma}, Z = {\rm
const.}
\end{equation}
After the application of the conformal time coordinate $\eta(t)=\int{dt/a(t)}$
the equations of motion and constraints
(\ref{ruch1})-(\ref{ruch3}) take the simple form
\bea
\label{eta1}
\ddot{\eta} &+& 2\frac{a_{,\eta}}{a} \dot{R}^2 = 0 ~,\\
\label{eta2}
\ddot{R} &+& 2\frac{a_{,\eta}}{a} \dot{\eta} \dot{R} + \lambda R =
0 ~, \\
\label{eta3}
\dot{\eta}^2 &-& \dot{R}^2 - \lambda R^2 = 0 ~.
\eea
The scale factor in phantom cosmology which admits a Big-Rip in conformal time scales as
\cite{phantom}
\begin{equation}
\label{solphantom}
a(\eta) = \eta^{-\frac{2}{3\mid \gamma \mid + 2}}~,
\end{equation}
where the barotropic index $\gamma = - \mid \gamma \mid < 0$ for phantom models
($p = (\gamma - 1) \varrho$, p - the pressure, $\varrho$ - the energy density).
In terms of the conformal time coordinate the invariant string
size is
\begin{equation}
\label{etasize}
S(\tau) = 2\pi a(\eta(\tau)) R(\tau)~.
\end{equation}
The simple solution of the system (\ref{eta1})-(\ref{eta3}) is \cite{FRWinigo}
\begin{equation}
\label{etaRtau}
\eta(\tau) = \exp{\left( \pm \sqrt{\frac{3 \mid \gamma \mid + 2}{2 - 3 \mid \gamma
\mid}} \tau \right)},  R(\tau) = \frac{\eta(\tau)}{2} \sqrt{3 \mid \gamma \mid +
2}  ~,
\end{equation}
so that the invariant string size reads as $(\mid \gamma \mid < 2/3)$
\begin{equation}
\label{Seta}
S(\tau) = \pi \sqrt{3 \mid \gamma \mid + 2} \times \exp{\left( \pm \frac{3 \mid \gamma
\mid}{\sqrt{4 - 9\gamma^2}} \tau \right) } ~.
\end{equation}
From (\ref{solphantom}) we see that a Big-Rip singularity $a(\eta) \to \infty$
appears for $\eta \to 0$. This singularity corresponds to the
limit $\tau \to \infty$ and the $`+'$ sign or $\tau \to - \infty$
and the $`-'$ sign in (\ref{etaRtau}). Then, it is clear from
(\ref{Seta}) that a string size is infinite in either of these
limits. This means a string will be {\it infinitely stretched} at a Big-Rip
singularity.

Now, let us discuss tensile strings at a sudden future singularity.
Let us choose the following evolution of the scale factor of an SFS model, which
presumably extends on both sides of sudden singularity, i.e.,
\begin{eqnarray}
\label{sfsleft}
a(t)&=& 1+\left(1 + \frac{t}{t_{B}} \right)^q(a_0 -1) - \left(\frac{-t}{t_{B}} \right)^n,
(t < 0)~~ \\
\label{sfsright}
\tilde{a}(t)&=& 1+\left(1 - {t \over t_{C}}\right)^q (a_0 -1) -
\left({t \over t_{C}}\right)^n,
(t > 0)~~
\end{eqnarray}
Here the sudden singularity appears at $t=0$, where $a_0 = a(0)=$ const., and
$0 < q \leq 1 $, $1 < n < 2$ \cite{barrow04,lake,barrow042,sfs1}.
The evolution begins with a Big-Bang singularity at $t=-t_{B}
<0$, faces a SFS at $t=0$, and finally reaches a Big-Crunch
singularity at $t=t_{C} >0$. We will show that the invariant
string size is finite at a sudden singularity. For simplicity,
we apply asymptotic solutions around $t=0$, i.e.,
\bea
\label{approx1}
a(t)&\approx& a_0 + \frac{q(a_0 - 1)}{t_{B}} t + \ldots, \hspace{0.1cm} (t <
0)~, \\
\label{approx2}
a(t)&\approx& a_0 - \frac{q(a_0 - 1)}{t_{C}} t + \ldots, \hspace{0.1cm} (t >
0)~.
\eea
Introducing the conformal time for (\ref{approx1})-(\ref{approx2})
we have
\begin{equation}
\eta = \frac{1}{\beta} \ln{\mid a_0 \pm \beta t \mid} ~,
\end{equation}
where $\beta=q(a_0 -1)/t_{B,C}$, and the scale factor is
\begin{equation}
\label{aeta}
a(\eta) = e^{\beta \eta}~.
\end{equation}
Then, one can see that at a SFS, $\eta$ is
finite which means that the scale factor (\ref{aeta}) is
finite and so the invariant string size at a SFS is finite, too. This
means a string may {\it cross smoothly} a SFS singularity and
eventually approach a Big Crunch singularity where it may
collapse to a zero size.

Let us now come to the discussion of the problem for the null strings
($\lambda=0$ in (\ref{ruch1})-(\ref{ruch3})).
We will briefly show that the null string can be considered as a
collection of particles in which $\sigma$ is assigned to a null
particle in a collection, while $\tau$ is a parameter of a geodesic
with a particular value of $\sigma$. Formally, $X^{\mu}(\tau,\sigma)$
is a geodesic with an index $\sigma$.

We note that the left-hand sides of the constraints
(\ref{ruch2}) and (\ref{ruch3}) are the constants of motion
for a collection of particles being the null string. The first
claim is trivial, while the second requires the discussion of an
absolute derivative of $g_{\mu\nu} \dot{X}^{\mu} {X'}^{\nu}$,
i.e.,
\bea
{d \over d\tau}(g_{\mu\nu} \dot{X}^{\mu} {X'}^{\nu}) &=& \dot{X}^{\rho}
\nabla_{\rho} (g_{\mu\nu} \dot{X}^{\mu}
{X'}^{\nu})=
\dot{X}_{\nu}\dot{X}^{\rho} \nabla_{\rho} {X'}^{\nu} \nonumber \\
&=& k_{\nu}((\pounds_{\vec{k}}
\vec{\eta})^{\nu} + \eta^{\rho} \nabla_{\rho} k^{\nu})\ ,
\eea
where $k^{\mu} = \dot{X}^{\mu}$ and ${\eta}^{\mu} = {X'}^{\mu}$.
Since $\vec{\eta}(\tau,\sigma) = \phi^*
\vec{\eta}(\tau_0,\sigma)$, where $p(\tau,\sigma)= \phi_{\tau}
p_0(\sigma)$ is a geodesic attached to an index $\sigma$, with
$p_0(\sigma)= \phi_{(\tau=\tau_0)} p_0(\sigma)$ ($\phi_{\tau}$ is
a map generated by a vector field $\vec{k}$), then
according to the properties of the Lie derivative the first term vanishes.
The second term vanishes due to the relation
\bea
k_{\nu} \eta^{\rho} \nabla_{\rho} k^{\nu}&=&\eta^{\rho} \nabla_{\rho} (k_{\nu}  k^{\nu})
- \eta^{\rho} k^{\nu} \nabla_{\rho} k_{\nu}
\nonumber \\
&=& - \eta^{\rho}  k^{\nu} \nabla_{\rho} k_{\nu}= - k_{\nu} \eta^{\rho} \nabla_{\rho}
k^{\nu} ~.
\eea
In conclusion, the evolution of the null string can {\it always be
reduced} to the evolution of the geodesics.

Let us now consider a circular string (\ref{circular}) which is equivalent to
a collection of particles with initial conditions $\varphi(\tau_0,\sigma) =
\sigma$, $\dot{\varphi}(\tau_0,\sigma)=0$. The first integrals of
(\ref{eom1})-(\ref{eom4}) for the null strings are
\begin{eqnarray}
\label{t}
\dot{t}^2&=&{A\over a^2(t)},\\
\label{r}
\dot{r}&=&{{B \sin \theta + P_3 \cos \theta} \over a^2(t)f(r)}, \\
\label{theta}
\dot{\theta}&=&{C\over a^2(t)r^2},\\
\label{varphi}
\dot{\varphi}&=& 0,
\end{eqnarray}
where $A=P^2+kL^2$, $B=P_1 \cos \sigma + P_2 \cos \sigma$, $C=L_1
\cos \sigma +L_2 \sin \sigma$ are constants independent of $\sigma$ \cite{lazkoz04}.
In terms of the cosmic time $t$ we calculate the coordinate velocity components as
\begin{eqnarray}
\label{vel1}
{dr\over dt}&=&{{B' \sin \theta + P_3' \cos \theta} \over a(t)f(r)}, \\
\label{vel2}
{d \theta \over dt}&=&{C'\over a(t)r^2},
\end{eqnarray}
where $B'=B/|A|^{1/2}$, $C'=C/|A|^{1/2}$,
$P_3'=P_3/|A|^{1/2}$.
Taking a horizontal plane $X^2\equiv \theta=\theta_0=$ const.
we formally impose an additional initial condition as
$\dot{\theta}(\tau_0,\sigma)=0$, which requires $C=0$.
For this particular choice of $\theta$ we have for
$k=1,0,-1$,
\begin{eqnarray}
r(t) = \sin\left(D\left(\int\limits_{0}^t{dx\over
a(x)}\right)\right), \\
r(t) = D\left(\int\limits_{0}^t{dx\over
a(x)}\right),\\
r(t) = \sinh\left(D\left(\int\limits_{0}^t{dx\over
a}\right)\right),
\end{eqnarray}
respectively, and $D=B'\sin \theta_0 + P_3' \cos \theta_0$.

Now let us come to the problem of the null strings at SFS.
Since for $t=t_s$ all the quantities in
the expression for the invariant string size (\ref{circsize}) are
finite, then {\it the size of the string at a sudden future
singularity is finite}. In all three cases $k=0,\pm 1$ the string
is spanned on the surface of either of the two cones of an angle $\theta_0$
each. The cones are attached to each other with their apexes at $r=0$.
The cones have a common symmetry axis given by $\theta=0$. For $k=+1$
the string oscillates around $r=0$ and the frequency of its
oscillations decreases while its invariant size (\ref{circsize}) grows,
though not to infinity.
For $k=0$ the string escapes from the point $r=0$ -- its
coordinate velocities (\ref{vel1})-(\ref{vel2}) decrease while its size grows.
For $k=-1$  the string escapes from $r=0$, its coordinate velocities and size grow
rapidly.

Now let us study the null strings at Big-Rip. We assume a simple model of evolution
of the scale factor which admits such a singularity
\cite{caldwell}, i.e.,
$a(t) = {a_R}(t_m)[\mid \gamma \mid + 1 - (\mid \gamma \mid + 1) ({t\over t_m})]^{-2/3\mid \gamma \mid},$
where $\gamma < 0$ and $t_m$ is the time of a Big-Rip.
The integral $\int{dx/a(x)}$ is convergent in an arbitrary small
interval before a BR singularity. For $k=+1$ the frequency of oscillations
of the string size is decreasing. Besides, the coordinate
velocities (\ref{vel1})-(\ref{vel2}) asymptotically tend to zero at a Big-Rip.
For $k=0$ the coordinate velocities (\ref{vel1})-(\ref{vel2}) decrease rapidly
in order to finally reach zero. Since $a(t) \to \infty$ at Big-Rip and all other quantities
in the invariant string size (\ref{circsize}) are finite, it means
that {\it the string will be infinitely stretched at a Big-Rip}, i.e., its size
$S \to \infty$. However, as concluded from the equations (\ref{t})-(\ref{varphi}),
in the limit $a(t) \to \infty$ all the right-hand sides of these geodesic
equations are zero which means that the four-velocity $dX^{\mu}/d\tau$ tends to zero
at a Big-Rip. This is not the case for the acceleration vector $d^2X^{\mu}/d\tau^2$
which, as seen from the geodesic equations (\ref{eom1})-(\ref{eom4}), is not regular
due to a blow-up of the scale factor $a(t)$ and its derivatives at
a Big-Rip.

Finally, we briefly discuss strings at some other types of future
singularities described in details in Refs. \cite{typeIII,typeIV}.
These singularities appear for the scale factor of the form
$a(t) = a_s \exp{\{h_0(t-t_s)^{1-\alpha}\}}$ and $h_0=$ const. If $0<\alpha<1$,
then $a=a_s=$ const., $\varrho \to \infty$, $\mid p \mid \to \infty$ at
$t=t_s$ (type III singularity of Ref. \cite{typeIII}). On the
other hand, if $\alpha < -1$, then $a=a_s$, $\varrho \to 0$ (or finite),
$\mid p \mid \to 0$ (or finite) and the higher order than two derivatives of the
scale factor diverge at $t=t_s$ (type IV singularity of Ref. \cite{typeIV}).
In the former case and for $k=+1$ the frequency of oscillations of a string approaching the
singularity grows though remains finite at $t=t_s$. What is
important from the point of view of the main task of our discussion is the fact that
the invariant string size (\ref{circsize}) remains finite at the
singularity. For $k=0,-1$ the coordinate velocities (\ref{vel1})-(\ref{vel2}) near $t=t_s$
grow rapidly. The integral $\int dx/a(x)$ is convergent on an
arbitrary interval of time before a singularity so that the
coordinate velocities and the radial coordinate $r$ are finite which
means that the invariant string size is also finite at $t=t_s$.
In the latter case for $k=+1$ one has almost homogeneous
and finite frequency of oscillations of a string near $t_s$. The invariant size
remains finite, too. For $k=0$ the coordinate velocities remain
almost constant near $t_s$ and the invariant size is finite.
For $k=-1$ one has a rapid growth of the coordinate velocities -
since $r$ remains finite, then the invariant size also remains
finite.

Not restricting the value of the azimuthal coordinate $\theta= \theta(\tau)$
we have
\begin{equation}
\label{elip}
{1\over {r^2}} = [Z \cos(\theta + \zeta)- G]^2 +k,
\end{equation} where $\zeta= \arctan {P'_3/B'}$ and $Z = B'/(C' \sin \zeta)$.
Since (\ref{elip}) restricts the value of $r$, then this $\theta= \theta(\tau)$ case
does not change a general picture in which the strings are not
infinitely stretched at these future singularities. Exact solutions for $\eta(\theta)$ can easily be
given but since they do not contribute a new quality to the
discussion we will not present them here.

In conclusion, we emphasize that bearing in mind the system of
geodesic equations (\ref{t})-(\ref{varphi}) one can see that the
geodesics may be extended through a sudden type, type III and type IV
singularities discussed
in this paper. This is a consequence of the finiteness of the
right-hand sides of these equations since in all the cases we have
studied the radial coordinate $r$ was finite which implied the
finiteness o the function $f(r)$. Then, strings may survive these
singularities. On the other hand, at a Big-Rip, geodesic equations
are singular (though the four-velocity is zero) and the strings
will be infinitely stretched. Finally, it is worth mentioning that
in Ref. \cite{tolley} the propagation of strings
through a Big-Crunch/Big-Bang singularity in the ekpyrotic/cyclic
scenario \cite{turok} was investigated, which is complementary
to our work about string propagation at Big-Rip and sudden future singularities.

\begin{center}
{\bf Acknowledgments.}
\end{center}

M.P.D. acknowledges the support of the Polish Ministry of
Education and Science grant No 1 P03B 043 29 (years 2005-2007).

\end{document}